\documentclass[ twocolumn,unsortedaddress,prl]{revtex4-1}
\usepackage{graphicx}
\usepackage{amsmath,amssymb}

\begin{document}


\title{The probability of double-strand breaks in giant DNA decreases markedly as the DNA concentration increases. }


\author{Shunsuke F. Shimobayashi}
\thanks{Email address}
\email{shimobayashi@chem.scphys.kyoto-u.ac.jp}
\affiliation{Department of Physics, Graduate School of Science, Kyoto University, Kyoto 606-8502, Japan}

\author{Takafumi Iwaki}

\affiliation{Fukui Institute for Fundamental Chemistry, Kyoto University, Kyoto 606-8103, Japan}

\author{Toshiaki Mori}
\affiliation{Radiation Research Center, Osaka Prefecture University, Sakai 599-8570, Japan}
\author{Kenichi Yoshikawa}
\thanks{Email address}
\email{yoshikaw@scphys.kyoto-u.ac.jp}
\affiliation{Department of Physics, Graduate School of Science, Kyoto University, Kyoto 606-8502, Japan}
\affiliation{ Faculty of Life and medical Sciences, Doshisha University, Kyoto 610-0394, Japan}


\date{\today}

\maketitle
\textbf{DNA double-strand breaks (DSBs) represent a serious source of damage for all living things and thus there have been many quantitative studies of DSBs both {\it in vivo} and {\it in vitro}~\cite{immunofluorescence, immunofluorescence2, a, b, c, ITO, d}. Despite this fact, the processes that lead to their production have not yet been clearly understood, and there is no established theory that can account for the statistics of their production, in particular, the number of DSBs per base pair per unit Gy, here denoted by $P_{1}$, which is the most important parameter for evaluating the degree of risk posed by DSBs. Here, using the single-molecule observation method with giant DNA molecules (166 kbp), we evaluate the number of DSBs caused by $\gamma$-ray irradiation. We find that $P_{1}$ is nearly inversely proportional to the DNA concentration above a certain threshold DNA concentration. A simple model that accounts for the marked decrease of $P_{1}$ shows that it is necessary to consider the characteristics of giant DNA molecules as semiflexible polymers to interpret the intrinsic mechanism of DSBs. }

As for DSBs caused by $\gamma$-rays {\it in vitro} experiments, its reported values scatter between $7.1\times10^{-9}$~\cite{a} and $2.1 \times10^{-6}$~\cite{c} (expressed as the number of DSBs per base pair per unit Gy). The scattered data may be attributed to significantly varying DNA concentrations and the sizes of DNA. 
Although gel electrophoresis has been the most widely used method to enumerate DSBs, it requires a high DNA concentration (more than several tens of $\mu$M in base pair units)~\cite{a, b, c, d, ITO}. Contrastingly, the single-molecule observation method employed in the present study is applicable to systems with DNA concentrations as small as three orders of magnitude smaller than those used in gel electrophoresis~\cite{ d}. In this article, we report the marked dependence of DSBs on the DNA concentration over a wide range of concentrations, indicating the significance for genome-sized DNA.

\begin{figure}[ht!bp]
 
 \centering
 \includegraphics[scale=0.5]{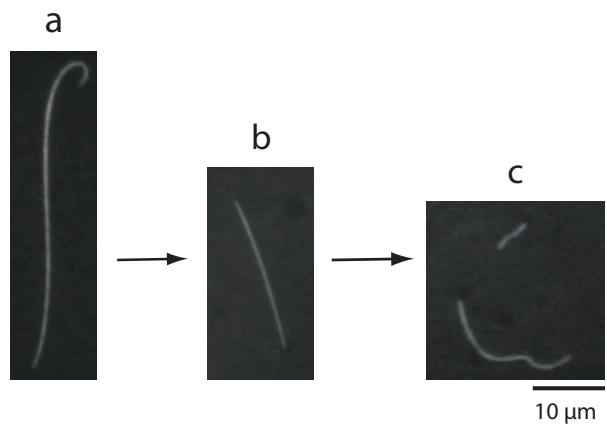}

 \caption{\label{image} {\bf Representative fluorescence images of DSBs for T4 DNA. }T4 phage 166kbp DNA in a 20 mM phosphate buffer solvent was irradiated by $\gamma $-rays from a {Co}$^{60}$ source. The fluorescent dye YoYo-1 was used for visualization of the DNA chains. DNA molecules in an aqueous solution were elongated through the effect of a weak shear applied by a glass slip and fixed on a poly-L-lysine coated glass. {\bf a}, 0 Gy (natural length). {\bf b}, 50 Gy. {\bf c}, 100 Gy.
}
\end{figure}

Figure \ref{image} exemplifies the fluorescence microscopic images that we obtained of the DNA in solution on the glass substrate. We obtained the average length of molecules irradiated with a dose $I$, $L(I)$, for a range of $I$ and thereby evaluated the number of DSBs per base pair, denoted by $P_{0}$ (see the Methods section).

\begin{figure}[htbp]
 \centering
 \includegraphics[scale=0.3]{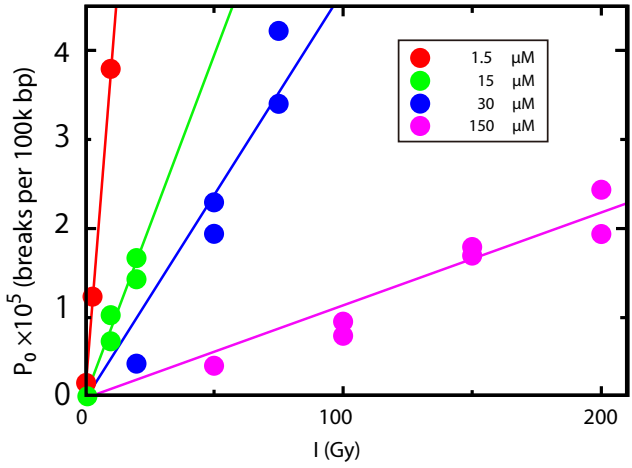}
 \caption{\label{linear} {\bf Irradiation dose dependence of the number of DSBs per base pair.} The number of DSBs per base pair, $P_{0}$, as a function of the irradiation dose from the {Co}$^{60}$ source, $I$. The data points for four different concentrations of samples are plotted. }

\end{figure}

\begin{figure}[h!]
 \centering
 \includegraphics[scale=0.3]{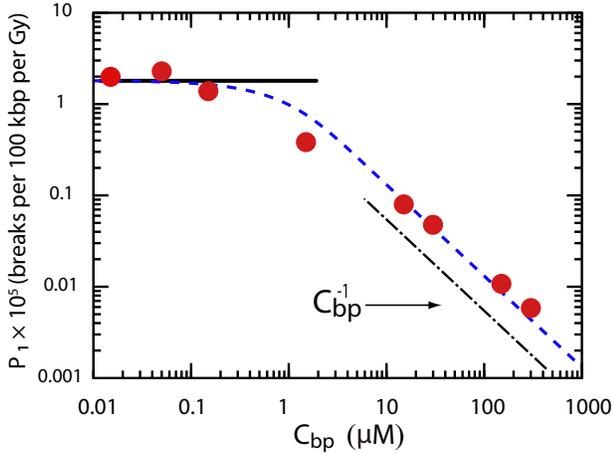}
 \caption{\label{fig:log-log} {\bf DNA base-pair concentration dependence of the number of DSBs per base pair per unit Gy.} Log-log plot of the number of DSBs per base pair per Gy, $P_{1}$, as a function of the DNA base-pair concentration, $C_\text{bp}$. It is seen that $P_{1}$ is roughly constant for small values of $C_\text{bp}$ (compare with the solid line of zero slope) and inversely proportional to $C_\text{bp}$ for large values. The blue dashed line represents a least-squares fit to the data using equation (\ref{e}). The best fit is obtained with $N_\text{rad}q/V_\text{tot}=1.31$ $\mu$M $(\text{100}\; \text{kbp} \cdot \text{Gy})^{-1}$  and $V/N_\text{bp}=1.38$ $(\mu\text{M})^{-1}$.
}
 
\end{figure}%

Figure \ref{linear} displays the dependence of $P_{0}$ on $I$. In the range of DNA concentrations from 0.015 to 300 $\mu$M in base pair units, $P_{0}$ for all samples showed linear dependences on $I$. Thus, $P_{1}$ for a given $C_\text{bp}$, the DNA base-pair concentration, is uniquely determined as the slope of the corresponding line representing $P_{0}(I)$.


Figure \ref{fig:log-log} contains a log-log plot of $P_{1}$ as a function of $C_\text{bp}$. Two different regimes are clearly seen. First, in the low concentration region, the quantity $P_{1}$ is nearly independent of $C_\text{bp}$. Second, in the high region, $P_{1}$ decreases as a function of $C_\text{bp}$, being approximately proportional to $C_\text{bp}^{-1}$. 
 

\begin{figure}[htbp]
 \centering
 \includegraphics[scale=0.3]{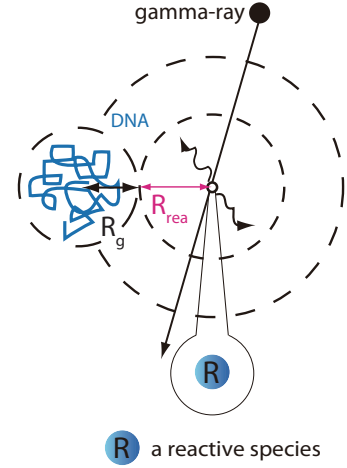}
\caption{\label{schematic} {\bf Schematic illustration of the interaction between a reactive species and DNA molecules.} We assume that a DSB is produced with a fixed probability $q$, provided that at least one DNA molecule exists in the sphere of radius $R_\text{rea}+R_\text{g}$. }
\end{figure}

\begin{figure}[htbp]
 \centering
 \includegraphics[scale=0.3]{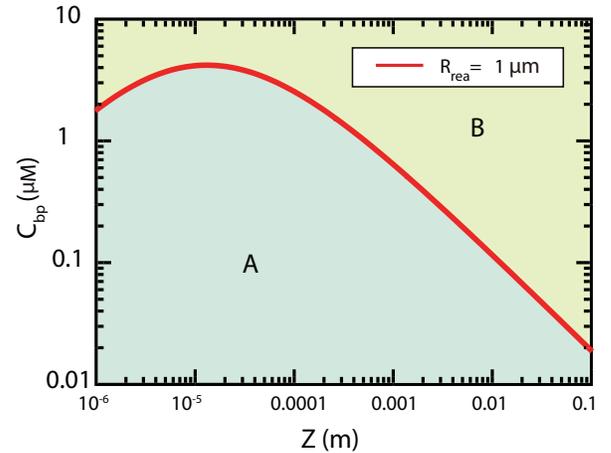}
\caption{\label{length} {\bf Theoretical prediction for the dependence of the threshold DNA base-pair concentration on the total length.} This phase diagram in the $Z$-$C_\text{bp}$ plane was obtained using equation (\ref{h}). Region A: $P_{1}$ is independent of $C_\text{bp}$. Region B: $P_{1}$ is dependent on $C_\text{bp}$.}
\end{figure}

The interaction of DNA molecules with ionizing radiation can be separated into
two categories~\cite{a sting, resonant formation, 0-4eV, bond breaks}, gdirect effects" (from energy deposited in the DNA and hydrated water molecules) and gindirect effects"  (from free radicals produced by energy deposited in water molecules and other biomolecules located close to the DNA). In dilute solution, the ratio of the energy deposited in DNA and hydrated water molecules would be much smaller than that deposited in the other molecules of the solution. In addition, Ito {\it et al.} has shown that the contribution of indirect effects to DSBs at $C_\text{bp}\cong 100$ $\mu \text{M}$ is approximately two orders of magnitude larger than that of direct effects~\cite{ITO}. Therefore it is reasonable to assume that indirect effects play a dominant role in the production of DSBs in our experiment.


Now we propose a simple model to account for our experimental data. In this model, it is assumed that each $\gamma$-ray photon deposits energy along the radiation track mainly through Compton scattering (see FIG.\ref{schematic})~\cite{compton} and produces reactive species~\cite{radiation}. Each reactive species has an effective diffusion length $R_\text{rea}$. According to a scaling argument, the radius of gyration $R_\text{g}$ can be regarded as defining the dimensions of a real DNA chain in a good solvent~\cite{de Gennes}. Given that at least one DNA molecule exists in the sphere of radius $R_\text{rea}+R_\text{g}$, we assume that a DSB is produced with a fixed probability $q$ (see FIG.\ref{schematic}). The sphere positioned at the center-of-mass of a reactive species means the gsphere of accessibility" for that reactive species : It is capable of attacking any DNA molecule located within in this sphere. Based on our framework, $P_{1}$ is given by
\begin{equation}
P_{1}=\frac{N_\text{rea}q(1-Q)}{N_\text{bp}N_\text{DNA}},
\label{d}
\end {equation}
where $N_\text{rea}$ is the number of reactive species produced per unit Gy, $Q$ is the probability that no DNA molecule exists in a given sphere, $N_\text{bp}$ is the number of base pairs for a single DNA molecule, and $N_\text{DNA}$ is the number of DNA molecules in the sample of total volume $V_\text{tot}$. The quantity $Q$ is obtained by assuming that a single DNA moelcule uniformly distributes in the region, that has the radius $R_\text{rea}+R_\text{g}$ and the volume $V$, in the sample of total volume $V_\text{tot}$ and that the positions of all DNA molecules are mutually independent variables. This yields the following:   
\begin{equation}
Q= (1-\frac{V}{V_\text{tot}})^{N_\text{DNA}}.   
\label{a}
\end {equation}
Next, because $N_\text{DNA}$ is given in terms of $C_\text{bp}$ as 
\begin{equation}
N_\text{DNA}=\frac{V_\text{tot}C_\text{bp}}{N_\text{bp}},
\label{b}
\end {equation}
for $V\ll V_\text{tot}$, we obtain
\begin{equation}
Q\approx e^{-\frac{V}{N_\text{bp}}C_\text{bp}}. 
\label{c}
\end {equation} 
Then, combining equations (\ref{d}), (\ref{b}) and (\ref{c}), we find
\begin{equation}
P_{1}\approx\frac{N_\text{rea}q(1-e^{-\frac{V}{N_\text{bp}}C_\text{bp}})}{V_\text{tot}C_\text{bp}}.
\label{e}
\end {equation}   
We have fitted our data with equation (\ref{e}) by treating $N_\text{rea}q/V_\text{tot}$ and $V/N_\text{bp}$ as fitting parameters (see FIG. (\ref{fig:log-log})). The best fit is obtained with $N_\text{rea}q/V_\text{tot}=1.31$ $\mu$M $(\text{100kbp} \cdot \text{Gy})^{-1}$ and $V/N_\text{bp}=1.38$ $(\mu\text{M})^{-1}$. 
In addition, $V/N_\text{bp}$ is given in terms of $R_\text{rea}$ and $R_\text{g}$ as 
\begin{equation} 
\frac{V}{N_\text{bp}}=\frac{\frac{4}{3}\pi (R_\text{rea}+R_\text{g})^{3}}{N_\text{bp}}.
\label{f} 
\end {equation}
Then, substituting the best-fit value for $V/N_\text{bp}$ and the values $N_\text{bp}=166\times 10^{3}$ and $R_\text{g}=1.5$ $\mu\text{m}$~\cite{Rg,Rg2}, we obtain $R_\text{rea}\sim 1 \mu \text{m}$ from equation (\ref{f}). Finally, the effective lifetime of the reactive species, $\tau$, is estimated to be on the order of one millisecond from the relation $R_\text{rea}^{2}\sim D\tau$, where $D$ $ (\sim10^{-9}$ $\text{m}^{2}/\text{s})$ is the diffusion coefficient~\cite{D}. This value of $\tau$ is reasonable, because the reactive species exist for approximately a millisecond~\cite{Ja}. 

The functional form of equation (\ref{e}) suggests that the threshold concentration $C_\text{bp}^{*}$ can be defined as the following relation:
\begin{equation} 
C_\text{bp}^{*}\sim  \frac{N_\text{bp}}{V}.
\label{g} 
\end {equation} 
The concentration $C_\text{bp}^{*}$ corresponds to the situation in which each region of volume $V$ contains one DNA molecule and each reactive species attacks just one DNA molecule, on average.
Therefore, the dependence in the high concentration region can be understood from the conjecture that the number of reactive species that can possibly attack a DNA molecule is essentially independent of $C_\text{bp}$, and thus increasing $C_\text{bp}$ results in the reduction of such reactive species per DNA molecule.

Furthermore, we take into account the size of DNA in order to study its dependence on $C_\text{bp}^{*}$. From equations (\ref{f}) and (\ref{g}) and the relations $N_\text{bp}=\frac{Z}{\eta }$ and $R_\text{g}\cong \frac{2}{3}\lambda _\text{k}^{2/5}Z^{3/5}$ ~\cite{de Gennes, Rg, Rg2}, $C_\text{bp}^{*}$ is obtained as follows:
 \begin{equation} 
   C_\text{bp}^{*}=\frac{Z}{\eta }\cdot\frac{1}{\frac{4}{3}\pi(R_\text{rea}+\frac{2}{3}\lambda _{k}^{2/5}Z^{3/5})^{3}},
   \label{h}
\end {equation} 
where $\eta$ (=3.4 \AA) is the distance between adjacent base pairs, $\lambda _\text{k}$ is the Kuhn length, and $Z$ is the total length of a single DNA molecule. The quantity $C_\text{bp}^{*}$ given by equation (\ref{h}) is plotted as a function of $Z$ in the case that $R_\text{rea}=1$ $\mu \text{m}$ and $\lambda _\text{k}=0.1$ $\mu  \text{m}$ in FIG.\ref{length}. This figure represents a phase diagram of $C_\text{bp}$ and $Z$: In region A, $P_{1}$ is independent of $C_\text{bp}$, while in region B, $P_{1}$ is inversely proportional to $C_\text{bp}$.

We now consider the process in which a single DSB is induced by reactive species. From the relations $P_{0} \propto  I $ (see FIG.\ref{linear}) and $I \propto  N_\text{pho}$ (the number of irradiating photons), we obtain the relation $P_{0} \propto  N_\text{pho} $. This assumes that DSB events produced by different photons are independent. Each gamma-photon produces clusters of reactive species along the radiation track~\cite{radiation, ms}. The diameters of these clusters are several nanometers, and they are widely separated in the case of radiation with low linear energy transfer~\cite{radiation, ms}. Futhermore, as shown in FIG.\ref{fig:log-log}, $P_{1}$ is inversely proportional to $C_\text{bp}$ above $C^{*}_\text{bp}$; i.e. the total number of DSBs is independent of $C_\text{bp}$. This implies that all reactive species attack DNA molecules and subsequently all of them disappear essentially. Therefore, this result indicates that the probability that a DSB will be produced is proportional to the number of reactive species that attack a single DNA molecule. With the scenario described to this point, we have two possibilities with regard to how a single DSB is produced: One possibility is that a single reactive species produces a single DSB with some probability, and the other is that a single cluster diffuses as an aggregate, attacks a single DNA molecule and produces a single DSB with some probability. However, the second possibility is seen to be unlikely, because the reactive species of each cluster diffuse almost homogeneously~\cite{ radiation, ms}. This argument thus leads to the conclusion that a single reactive species induces a single DSB with some probability (see FIG.\ref{schematic}). Our model, which is based on this scenario, yields results consistent with our experimental data(see FIG.\ref{fig:log-log}).

In conclusion, employing the single-molecule observation method, we find that $P_{0}$ is linearly dependent on $I$ less than 200Gy, and that $P_{1}$ remains nearly constant below a threshold value $C_\text{bp}^{*}$ and $P_{1}$ is nearly inversely proportional to $C_\text{bp}$ above $C_\text{bp}^{*}$. Our theoretical model indicates that $P_{1}$, the number of DSBs per base pair per unit Gy, decrease as the DNA concentration increases above a certain threshold because the number of potentially attacking reactive species per DNA molecule also decreases. We conclude that it is necessary to consider not only the microscopic properties of the DNA molecule, but also the characteristics of giant DNA molecules as semiflexible polymers and the density characterizing the entire collection of such molecules for the systems in which they exist. This is a significant finding in our endeavor to understand the physics of radiation-induced ionization of DNA molecules, and it also has important implications for biology and medicine.

\section{methods}
{\bf Experimental procedures and evaluation method.} We used T4 phage DNA (166kbp, Nippon Gene) in a 20 mM phosphate buffer (pH 7.3) solvent. DNA was irradiated by $\gamma$-rays from a {Co}$^{60}$ source at a dose rate of 0.002-7 Gy/min. The fluorescent dye YO-YO-1 (Abs/Em 491/509 nm, Molecular Probes, USA)  was used for visualization of the DNA chains. After the irradiation, we added YO-YO-1 at $1\mu$M and 2-mercaptoethanol (Nacalai Tesque, Japan) at 4\% (v/v) at the final concentration in order to avoid additional oxidative damage. To fix the DNA molecules on a glass substrate, a glass base dish of 35 mm diameter (Iwaki, Japan) was cationically modified with a 0.05\% (v/v) poly-L-lysine solution (Sigma-Aldrich, USA), and a sample solution ($\sim$5 $\mu$l) was adsorbed onto the modified glass base dish. To elongate the DNA molecules, the droplet of sample solution was covered with a glass slip (18 mm$\times$18 mm). We applied a weak shear to the DNA molecules by sliding this glass slip. This was done very gently to avoid additional fragmentations by mechanical stress. Fluorescence images of DNA molecules were observed using an Axiovert 135 TV (Carl Zeiss, Germany) fluorescence microscope equipped with a 100$\times$ oil-immersed objective lens. The lengths of the elongated DNA molecules were measured using cosmos image-analysis software (Library, Japan). We measured the contour lengths of approximately 100 molecules in each sample solution and thereby obtained $L(I)$. Giant DNA molecules exceeding 100kbp are occasionally inadvertently cut into fragments during the sample preparation, which includes pipetting and mixing~\cite{mixing}. For this reason, the value of $L(0)$ is slightly shorter than the natural length of T4 DNA, 57 $\mu$m~\cite{compaction, damian}.
 
 The quantity $P_{0}$ (the number of DSBs per base pair) is calculated from the difference between the average molecular lengths of irradiated and non-irradiated samples that are otherwise identically prepared:  

\begin{equation}
P_{0}= \frac{L(0)/L(I)-1}{L(0)}  \cdot \eta .
\label{eq:P} 
\end{equation}
Here, $\eta$ (=3.4 \AA) is the distance between adjacent base pairs.





\section{acknowledgments}

We thank M. Ichikawa, Y. Maeda, Y. Yoshikawa, M. Suzuki, S. N. Watanabe S. Hirota, and M. Miyaji for useful discussions. This work was supported by a Sasagawa Scientific Research (Grant No. 24-621) from the Japan Science Society, and partly by Grant-in-Aid for Scientific Research (A) (No. 23240044) from the Japan Society for the Promotion of Science (JSPS).


\end{document}